 \documentclass[]{aa}
 \usepackage{graphicx}
 \usepackage{txfonts}
 \usepackage{tabularx}
 \newcommand{\sd}{SDSS~J0011$+$0055}
 \newcommand{\civ}{\ion{C}{iv}}
 \newcommand{\siiv}{\ion{Si}{iv}}
 \newcommand{\mgii}{\ion{Mg}{ii}}
 \newcommand{\nv}{\ion{N}{v}}
 \newcommand{\aliii}{\ion{Al}{iii}}
 \newcommand{\cii}{\ion{C}{ii}}
 \newcommand{\siii}{\ion{Si}{ii}}
 \newcommand{\ciie}{\ion{C}{ii}$^{\star}$}
 \newcommand{\siiie}{\ion{Si}{ii}$^{\star}$}
 \newcommand{\feii}{\ion{Fe}{ii}}
 \newcommand{\hi}{\ion{H}{i}}
 \newcommand{\ks}{km~s$^{-1}$}
 
\begin{document}
    \title{VLT + UVES Spectroscopy of the Low-Ionization
	Intrinsic Absorber in SDSS~J001130.56$+$005550.7
\thanks{Based on observations
     collected at the European Southern Observatory, Cerro Paranal, Chile 
     (ESO 267.A-5698)}}


    \author{Damien Hutsem\'ekers \inst{1,2}\thanks{Research Associate
            FNRS (Belgium)}, Patrick B. Hall  \inst{3,4},
            J. Brinkmann \inst{5}
}

    \institute{Institut d'Astrophysique, Universit\'e de Li\`ege,
               All\'ee du 6 Ao\^ut 17, Bat. B5c, B-4000 Li\`ege,
               Belgium \and European Southern Observatory, Casilla
               19001, Santiago 19, Chile \and Princeton University
               Observatory, Princeton, NJ, USA 08544 \and Departamento de
               Astronom{\'\i}a y Astrof{\'\i}sica, Pontificia Universidad
               Cat\'olica de Chile, Casilla 306, Santiago 22, Chile \and Apache
               Point Observatory, P.O. Box 59, Sunspot, NM, USA 88349-0059}

    \date{Received: 7 May 2003 ;  accepted: 31 October 2003}

\abstract{We analyse high-resolution VLT+UVES spectra of the
low-ionization intrinsic absorber observed in the BAL QSO
SDSS~J001130.56$+$005550.7.  Two narrow absorption systems at
velocities $-$600 km s$^{-1}$ and $-$22000 km s$^{-1}$ are
detected. The low-velocity system is part of the broad absorption line
(BAL), while the high-velocity one is well detached.  While most
narrow absorption components are only detected in the high-ionization
species, the lowest velocity component is detected in both high- and
low-ionization species, including in the excited \siiie\ and \ciie\
lines.
From the analysis of doublet lines, we find that the narrow absorption
lines at the low-velocity end of the BAL trough are completely
saturated but do not reach zero flux, their profiles being dominated
by a velocity-dependent covering factor. The covering factor is
significantly smaller for \mgii\ than for \siiv\ and \nv , which
demonstrates the intrinsic nature of absorber.
From the analysis of the excited \siiie\ and \ciie\ lines in the
lowest velocity component, we find an electron density $\simeq$
10$^{3}$ cm$^{-3}$. Assuming photoionization equilibrium, we derive a
distance $\simeq$ 20 kpc between the low-ionization region and the
quasar core.  The correspondence in velocity of the high- and
low-ionization features suggests that all these species must be
closely associated, hence formed at the same distance of $\sim$ 20
kpc, much higher than the distance usually assumed for BAL absorbers.
     }

    \titlerunning{VLT+UVES spectroscopy of SDSS~J001130.56$+$005550.7 }
    \authorrunning{Hutsem\'ekers et al.}

    \maketitle

\section{Introduction}

Intrinsic absorption lines in quasars are usually classified as broad
absorption lines (BALs) or narrow absorption lines (NALs). These
absorption line systems are to be distinguished from cosmologically
``intervening'' systems unrelated to the quasar environment (Barlow et
al. \cite{bar1997b}).

Broad (velocity width FWHM $>$ 2000 \ks ) troughs (BALs) are detected
in roughly 15\% of optically selected quasars (Hewett \& Foltz
\cite{hf2003}, Reichard et al.  \cite{rei2003b}).  They are
blueshifted with respect to the QSO emission lines.  BAL outflows
occur at velocities of typically 0.1$c$ (Weymann et
al. \cite{wey1991}). Most BAL QSOs have absorption in high-ionization
species like \civ\ $\lambda$1549, \siiv\ $\lambda$1397 and \nv\
$\lambda$1240. A minority of them also show absorption due to lower
ionization species (LoBAL) such as \mgii\ $\lambda$2798 or \aliii\
$\lambda$1857.

NALs have velocity widths of at most a few hundred \ks . NALs are not
only observed at redshifts $z_{\rm abs} \simeq z_{\rm em}$ but also
at blueshifted velocities comparable to those seen in BAL QSOs
(Barlow et al. \cite{bar1997b}, Hamann et al. \cite{ham1997a}). NALs
can also appear redshifted up to $\sim$ 2000 \ks even though they are
frequently blueshifted.  Since common doublet transitions are
resolved in NALs using high-resolution spectroscopy, they constitute
useful diagnostics of the quasar environment.  The so-called
mini-BALs (e.g. Churchill et al. \cite{chu1999}) have intermediate
absorption widths, i.e. FWHM between a few hundred and 2000 \ks .

BALs and NALs indicate that outflows from Active Galactic Nuclei span
a large range of velocity widths.  While it is known that NALs may
form in various environments (Hamann et al. \cite{ham2001} and
references therein), it is not clear whether at least some of them are
directly related to the BAL phenomenon. In some rare cases, NALs and
BALs may be observed in the same quasar, providing an opportunity to
directly investigate this issue.

In this paper we report high-resolution spectroscopy of
SDSS~J001130.56$+$005550.7 (hereafter \sd ; Schneider et al.
\cite{sch2002}) discovered in the Sloan Digital Sky Survey (York et
al. \cite{yor2000}).  This object is one of the $\sim$10$^5$ quasar
candidates (Richards et al. \cite{ric2002}) for which the survey is
obtaining redshifts, in addition to the $\sim10^6$ galaxies which
comprise the bulk of the spectroscopic targets (Blanton et
al. \cite{bla2003}), selected from astrometrically calibrated
drift-scanned imaging data (Gunn et al. \cite{gun1998}, Pier et
al. \cite{pie2003}) on the SDSS $ugriz$ AB asinh magnitude system
(Fukugita et al. \cite{fuk1996}, Lupton et al. \cite{lgs1999}, Hogg et
al. \cite{hog2001}, Stoughton et al. \cite{sto2002}, Smith et
al. \cite{smi2002}).

\begin{figure}
\resizebox{\hsize}{!}{\includegraphics*{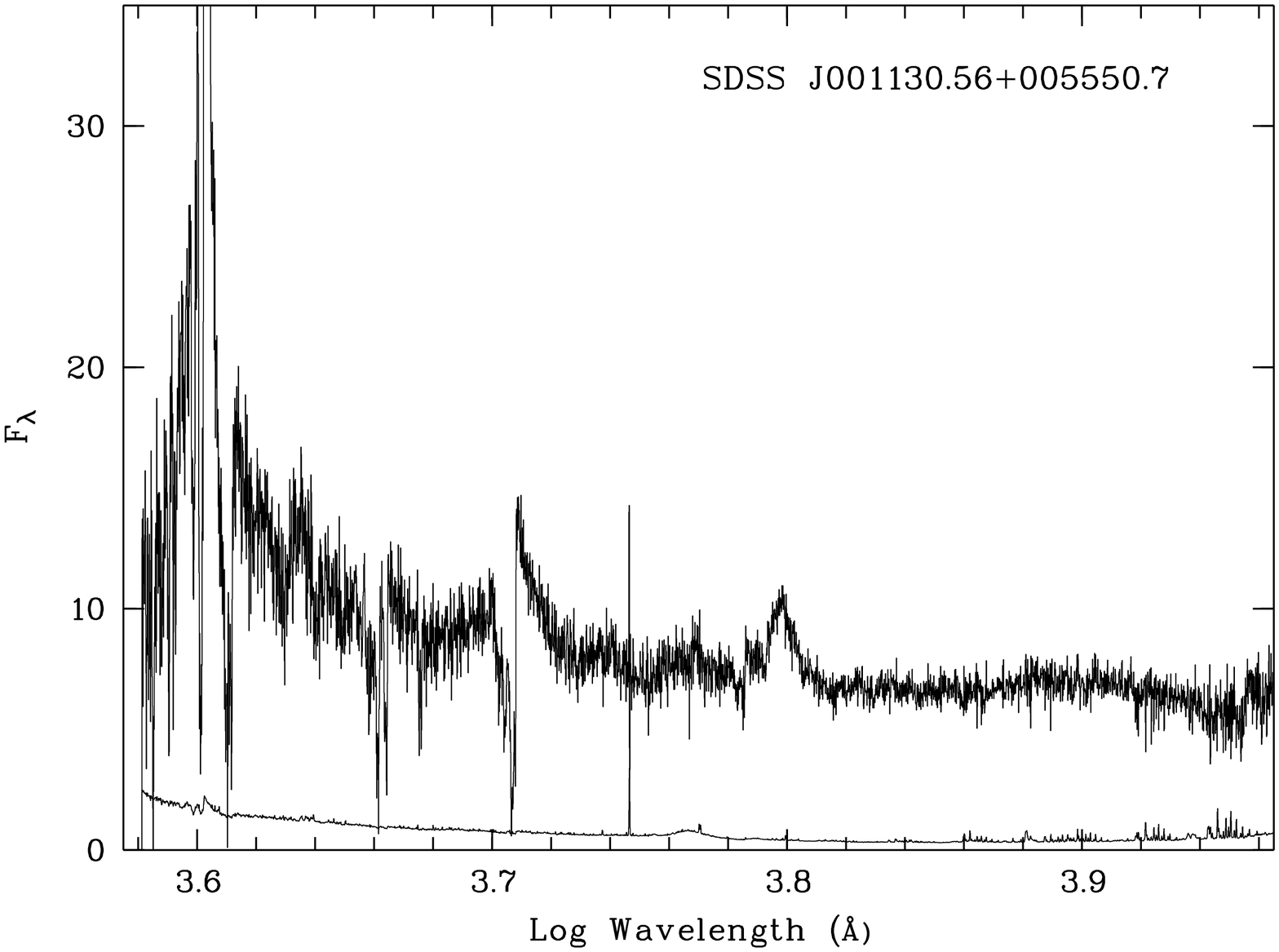}}
\resizebox{\hsize}{!}{\includegraphics*{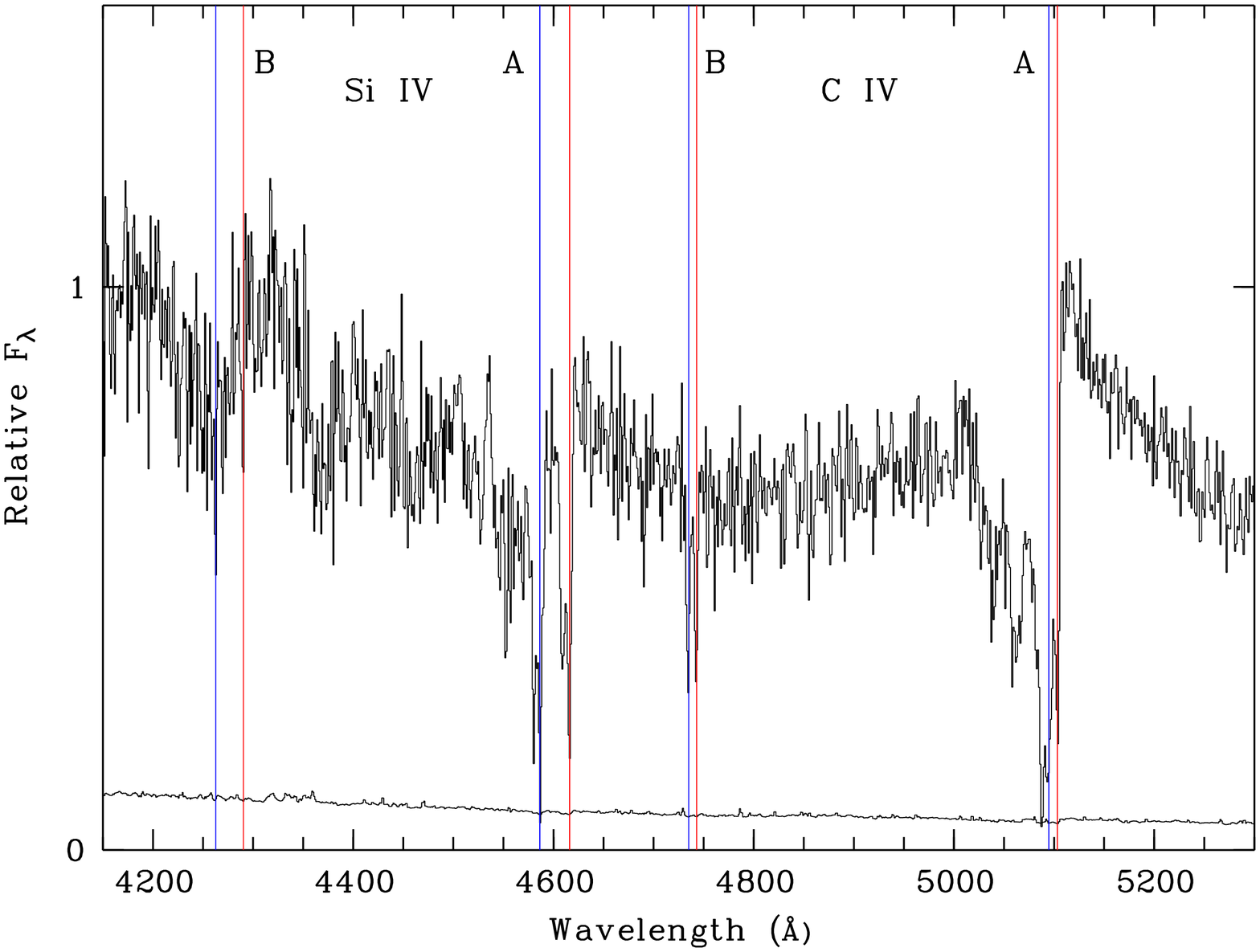}}
\caption{SDSS spectrum ($R \simeq 2000$) of \sd .Top: the full
 spectrum. $F_{\lambda}$ is in units of 10$^{-17}$ ergs cm$^{-2}$
 s$^{-1}$ \AA$^{-1}$. Bottom: a portion of the spectrum illustrating
 the \siiv\ $\lambda\lambda$ 1393.75, 1402.77 and \civ\
 $\lambda\lambda$ 1548.20, 1550.78 doublets. Absorption lines
 at $z$ = 2.29071 (part of system A) and at $z$ = 2.05844 (part of
 system B) are indicated}
\label{fig:sdss}
\end{figure}

\sd\ has strong intrinsic absorption extending up to a velocity of
$\sim$5500 \ks\ in \civ .  It just misses formal classification as a
BAL QSO: the balnicity index defined by Weymann et
al. (\cite{wey1991}) is measured to be zero using the method of
Reichard et al.  (\cite{rei2003a}).  However, it has a value of 2040
\ks\ on the absorption index scale designed by Hall et
al. (\cite{hal2002}) to include troughs too narrow or close to the
quasar redshift for consideration by the balnicity index.  Since the
formal classification is rather arbitrary and non-physical, we
consider \sd\ as a BAL QSO in the following.  

\sd\ shows both high- and low-ionization absorptions resolved into
narrow components, together with a well-detached high-velocity $z_{\rm
abs} \ll z_{\rm em}$ narrow line system. Because some absorption
arises from excited levels, \sd\ also provides a rare opportunity to
derive the electron gas density and to constrain the distance to the
absorber.

Observations are reported in Sect. \ref{sec:Obs} and the
characteristics of the spectrum in Sect. \ref{sec:Over}. The narrow
lines are analysed in Sect. \ref{sec:Ana} in order to establish their
intrinsic nature, to evaluate the covering factor of the different
ions, and to estimate the electron density in the flow using excited
lines. Discussion and conclusions form the last section.

 \section{UVES observations} 
 \label{sec:Obs}

Observations of several SDSS BAL quasars were obtained on UT
10-12 Aug 2001 using the ESO Very Large Telescope (VLT) Unit 2
(Kueyen) and the Ultraviolet-Visual Echelle Spectrograph (UVES).  Part
of these observations are reported in Hall et al. (\cite{hal2003}).

For \sd\ two hour-long exposures were secured in the UVES DIC2 437+860
standard setting (Kaufer et al. \cite{kau2001}) with the depolarizer
inserted.  A 1$\arcsec$ slit and a 2$\times$2 CCD binning were used,
yielding an overall spectral resolution $R \simeq 40000$ (7.5 \ks ).
Taking into account the fact that some orders are not useful due to
bad signal to noise and/or strong artifacts, good quality spectra were
obtained in the spectral ranges $\lambda\lambda$ 3760--4980~\AA ,
$\lambda\lambda$ 6700--8510~\AA\ and $\lambda\lambda$ 8660--10420~\AA
.

Each exposure was reduced individually using the dedicated UVES
pipeline (Ballester et al. \cite{bal2000}) developed within the ESO
Munich Image Data Analysis System (MIDAS).  Optimal extraction of the
spectra was performed, including simultaneous rejection of cosmic ray
hits and subtraction of the sky spectrum.  Telluric absorption lines
were removed for the red setting with the use of observations of
telluric standard stars, shifted in velocity according to the
different times of the observations and scaled in intensity according
to the airmass difference.  Additional cosmic ray rejection was done
by a detailed comparison of the two exposures, before co-addition and
merging.  The final 1-D spectrum was rebinned on a vacuum heliocentric
scale.

 \section{Overview of the spectrum} 
 \label{sec:Over}

\sd\ is a high redshift ($z_{\rm em}\simeq 2.3$) low-ionization BAL
QSO. Since only the rest-frame $\lambda\lambda$ 1140--1510 \AA ,
$\lambda\lambda$ 2040--2590 \AA\ and $\lambda\lambda$ 2630--3160 \AA\ 
are covered by UVES spectra, \civ\ $\lambda$1549 is not observed. We
show in Fig.~\ref{fig:sdss} a portion of the SDSS spectrum
illustrating this spectral region (Schneider et al.  \cite{sch2002}).

From the peak of the \mgii\ emission line at $\lambda$ 2803~\AA\ in
the UVES spectrum, we adopt a systemic redshift of $z_{\rm sys}$ =
2.29263.  This is slightly lower than the SDSS redshift of $z=2.30576$
(Schneider et al. \cite{sch2002}), but the SDSS spectrum does not
include the \mgii\ emission line.  The exact value does not
particularly matter, since all our discussions are in velocity space.

\begin{table}[tb]
\caption{The absorption systems studied in this paper}
\begin{tabular}{lcr}
\hline
 System & $z_{\rm abs}$     & FWHM \\
\hline
 A1 & 2.29071 $\pm$ 0.00004 & 174  \\
 A2 & 2.28724 $\pm$ 0.00002 &  27  \\
 A3 & 2.28626 $\pm$ 0.00003 &  62  \\
 B1 & 2.05844 $\pm$ 0.00002 & 104  \\
 B2 & 2.05740 $\pm$ 0.00003 &  24  \\
\hline
\end{tabular}
\label{tab:zabs}
\end{table}

Several narrow absorption line systems may be identified in the UVES
spectrum of \sd .  The characteristics of the systems studied in the
present paper are reported in Table~\ref{tab:zabs}. FWHM (in \ks ) are
measured from the \siiv\ and \civ\ lines for which these absorption
lines are best seen.  System~A consists of a cluster of resolved
narrow lines at the low-velocity end of the BAL trough seen in
the SDSS low resolution spectrum (Fig.~\ref{fig:sdss}). It is well
defined and resolved in \siiv\ as well as in \mgii , making this
system well suited for a detailed analysis.  The higher velocity
components seen in the \civ\ BAL trough (Fig.~\ref{fig:sdss}) appear
broad, shallow or blended in the UVES spectrum of \siiv\ and \nv\ and
undetected in \mgii\ such that they are not further considered in the
analysis.  While system~A is part of the BAL and superimposed on the
broad emission, system~B is a high-velocity narrow absorption system
well detached from the BAL trough and the broad emission
(Fig.~\ref{fig:sdss}). Additional (FWHM $\simeq 18$ \ks )
intervening systems are also detected at $z$ = 1.77889 in \civ,
\siiv\ and \mgii , at $z$ = 1.77791 in \civ , and at $z$ = 0.48727 in
\mgii .

\begin{figure*}[!ht]
\resizebox{\hsize}{!}{\includegraphics*{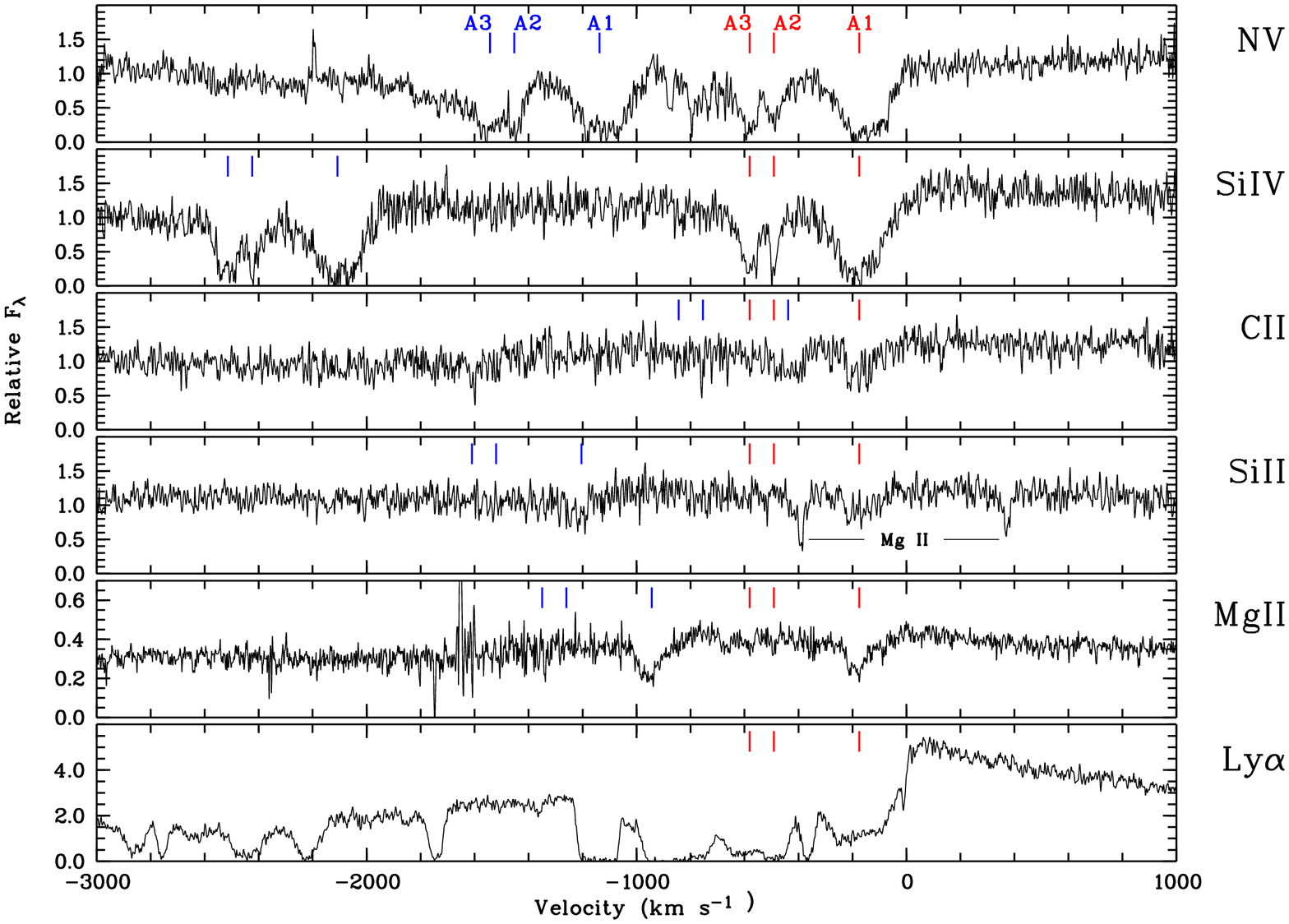}}
\resizebox{\hsize}{!}{\includegraphics*{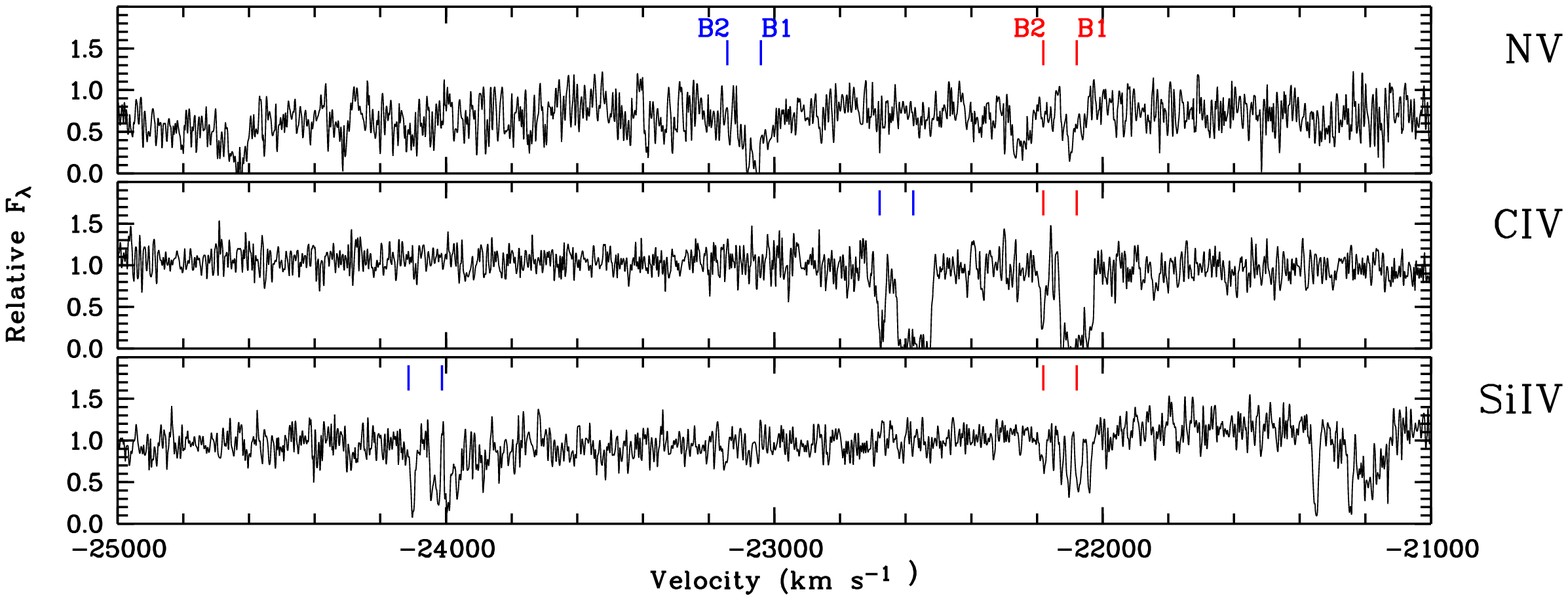}}
\caption{Portions of the UVES spectrum of \sd\ illustrating the
absorption line systems A and B defined in Table~\ref{tab:zabs}.
System B is best seen in \civ\ and system A in \siiv\ (\civ\ at $z
\simeq$ 2.29 is outside the observed UVES spectral range).  The
following transitions are indicated: \nv\ $\lambda\lambda$ 1238.82,
1242.80, \civ\ $\lambda\lambda$ 1548.20, 1550.78, \siiv\
$\lambda\lambda$ 1393.75, 1402.77, \mgii\ $\lambda\lambda$ 2796.35,
2803.53 , \siii\ $\lambda\lambda$ 1260.42, 1264.74, 1265.00, \cii\
$\lambda\lambda$ 1334.53, 1335.66, 1335.71, and Ly$\alpha$ $\lambda$
1215.67.  For clarity, the two reddest and very close transitions of
\siii\ and \cii\ are represented by single marks at the
mean wavelengths; all four of those transitions are excited-state
transitions.  Absorption from the high-velocity system B is not
detected in \siii\ or \cii , and therefore is not plotted although it
lies well within the observed UVES spectral range (\siii\ $\lambda$
1260.42 could be present in system B1 but the detection is not
convincing).  Narrow intervening \mgii\ $\lambda\lambda$
2796.35, 2803.53 absorption contaminates the \siii\ absorption of
system A2.  Narrow intervening \civ\ $\lambda$1548.20 lines are seen
slightly longward of the \siiv\ absorption from system B, along with a
broader instrumental artifact.  All spectra are plotted on a velocity
scale of $\rm{v}/c$ = $(R^{2}-1)/(R^{2}+1)$ where $R$ = $(1+z_{\rm
sys})\: \lambda_{0} / \lambda_{\rm obs}$, $\lambda_{\rm obs}$ being
the observed wavelength, $\lambda_{0}$ the laboratory vacuum rest
wavelength of the red transition of the multiplets, and $z_{sys}$ the
systemic~redshift~of~the~quasar.}
\label{fig:vel}
\end{figure*}

Portions of the UVES spectrum are illustrated in Fig.~\ref{fig:vel}.
The absorption system A1 is seen in the high- and low-ionization
species \nv , \siiv , \mgii , \siii , \cii\ and \hi , including
\siiie\ and \ciie\ excited states. No \feii\ absorption is detected. 
The velocity correspondence indicates that all the observed species
must be physically associated.  Component A1 is definitely broader
than the typical thermal velocity width ($\lesssim$ 10 \ks , Barlow \&
Sargent \cite{bar1997a}). No ionization dependent velocity
stratification can be observed in this component.  Different behavior
is seen in components A2 and A3, which are not detected in the
low-ionization species but only in \nv , \siiv , and Ly$\alpha$.  It
is worth emphasizing that broad absorption at higher velocity is seen
in \civ , \siiv\ and \nv\ while only the narrow component A1 is
clearly detected in \mgii\ in agreement with the fact that
low-ionization features are more often found at the low-velocity ends
of BAL troughs (Voit et al. \cite{voi1993}). Note that the important
diagnostic line \ion{Mg}{i} $\lambda$ 2853 is not detected in \sd.

The other interesting feature is the detached narrow absorption system
with a velocity $>$ 20000 \ks . This high-velocity system is detected
in the high-ionization species only and not in the low-ionization ones
(i.e. not in \siii\ nor \cii ; \mgii\ and Ly$\alpha$ at $z \simeq$
2.05 are not in the observed UVES spectral range).  It is relatively
narrow and clearly separated from the BAL trough, which reaches only
$\sim 5500$ \ks\ (Fig.~\ref{fig:sdss}). Two major components (named B1
and B2) are identified in the \civ\ line, while several narrower (15
-- 20 \ks\ FWHM) components form the unsaturated \siiv\ line. The fact
that component B1 is detected in \nv\ is suggestive of an intrinsic
origin.

 \section{Analysis of the spectrum} 
 \label{sec:Ana}

\subsection{Partial line-of-sight covering}

The ability to measure unblended features from two lines of the same
ion allows us to solve separately for the effective covering factor
and real optical depth, and to establish the intrinsic nature of the
absorbers (e.g. Barlow \& Sargent \cite{bar1997a}).

If the absorption region covers a fraction $C_{\rm v}$ of the quasar
light with an optical depth $\tau_{\rm v}$, then for 
unblended doublet absorption lines we have (e.g. Hall et al. 2003):
\begin{eqnarray}
I_{b} & = & 1 - C_{\rm v} ( 1 - e^{-\tau_{\rm v}}) \nonumber \\
I_{r} & = & 1 - C_{\rm v} ( 1 - e^{-0.5\, \tau_{\rm v}})         
\end{eqnarray}
from which we derive
\begin{equation}
C_{\rm v} =  \frac{1+I_{r}^{2}-2\:I_{r}}{1+I_{b}-2\:I_{r}} \\ \rm{and} \\
e^{-\tau_{\rm v}}  = 1-\frac{1-I_{b}}{C_{\rm v}} , 
\end{equation}
where $I_{b}$ and $I_{r}$ are the normalized residual intensities in
the blue and red lines of the doublet (when the blue line is also the
strongest). The solution for $C_{\rm v}$ --and $e^{-\tau_{\rm v}}$--
is only physical ($0 \leq C_{\rm v} \leq 1)$ when $0 \leq I_r^2 \leq
I_b \leq I_r \leq 1$.  When $I_{r}^{2} = I_{b}$, $C_{\rm v} = 1$ and
the line profile is determined solely by the opacity.  $I_{r}^{2} <
I_{b}$ implies $C_{\rm v} < 1$. When $I_{r} = I_{b}$, $I_r = 1-C_{\rm
v}$ and the line profile is saturated and essentially determined by
the covering factor.

This analysis requires the residual intensities to be normalized to
the underlying continuum.  Since the absorption lines at $z \simeq$
2.29 deeply cut the broad emission, we assume that the absorption also
covers the emission, and we adopt a local continuum that includes the
broad emission. We will see below that this hypothesis is adequate.

\subsubsection{The low-velocity absorption system}
 \label{sec:low}

\begin{figure}
\resizebox{\hsize}{!}{\includegraphics*{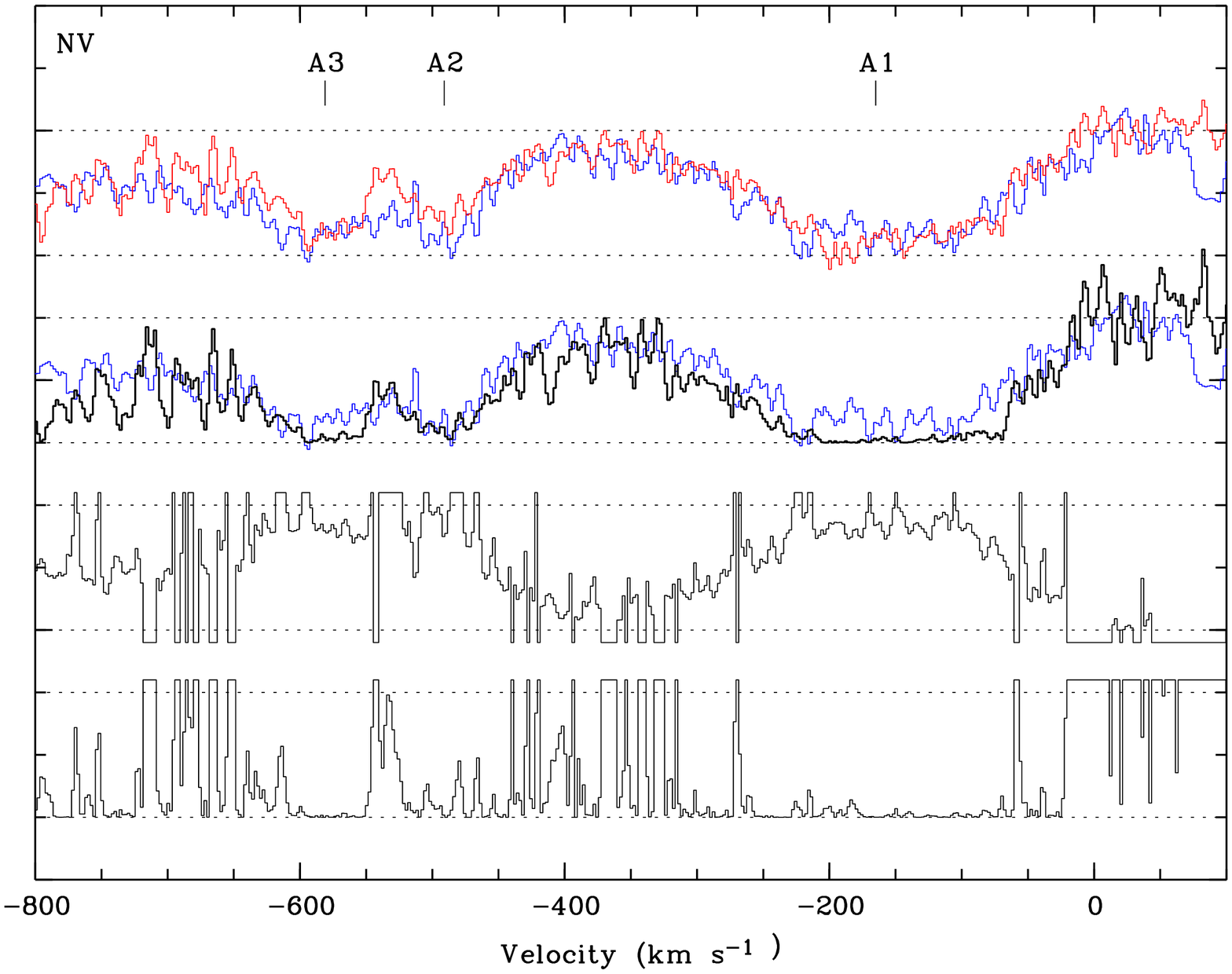}}
\resizebox{\hsize}{!}{\includegraphics*{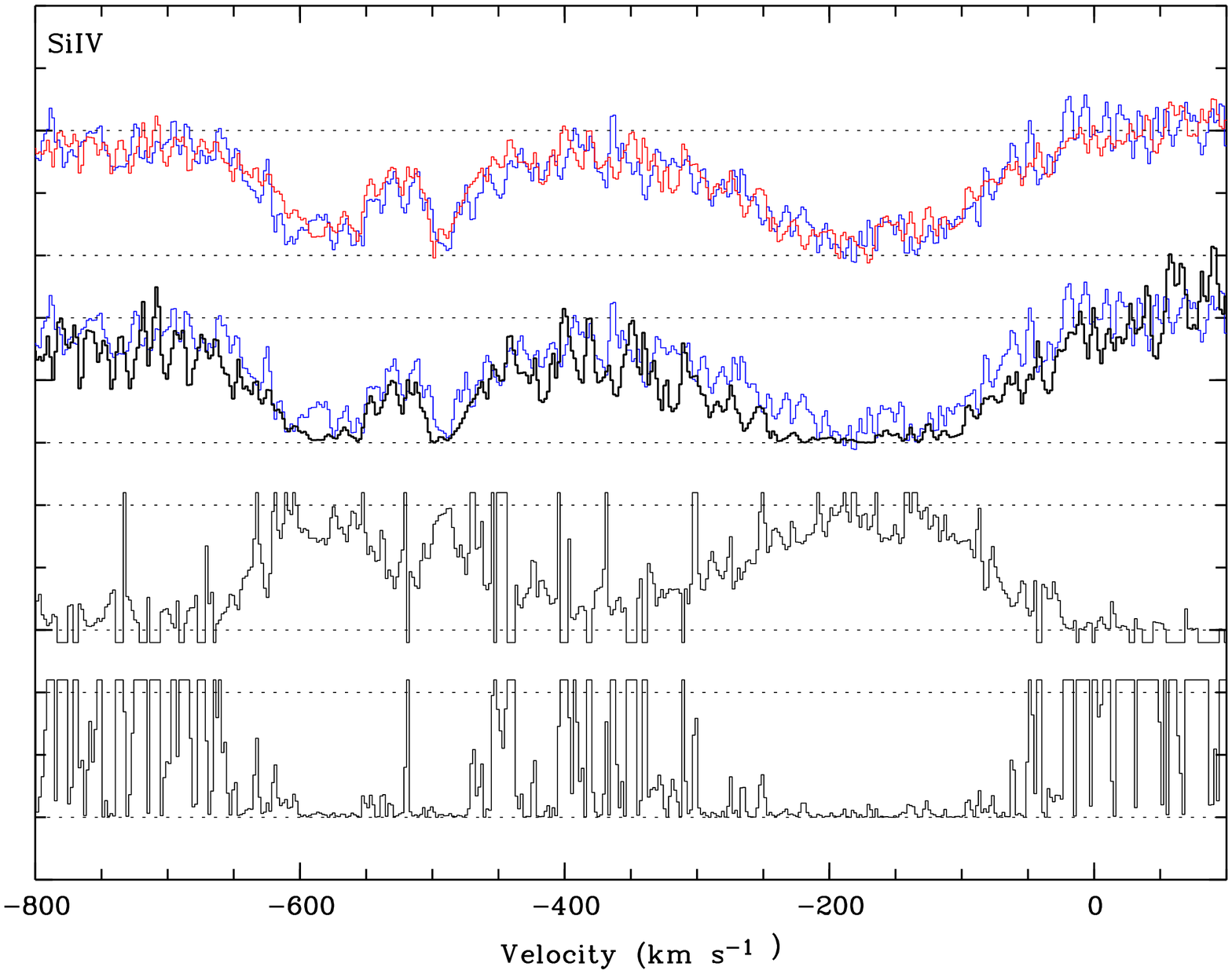}}
\resizebox{\hsize}{!}{\includegraphics*{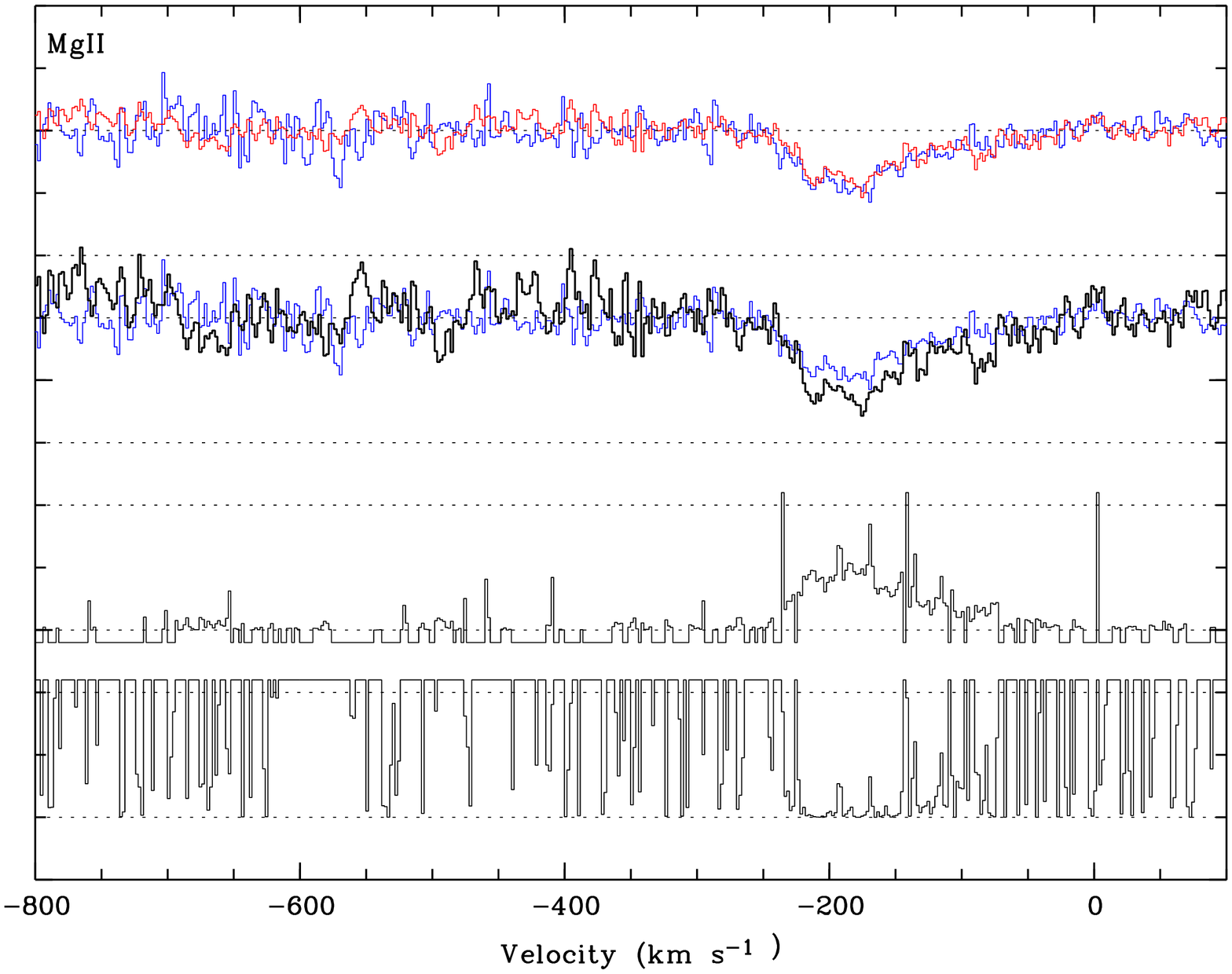}}
\caption{The covering factor and opacity of resonance doublets seen in
the low-velocity absorption system A and computed from Eq.~(2).  From
top to bottom: the normalized residual intensities $I_{b}$ (blue) and
$I_{r}$ (red); the normalized residual intensity $I_{b}$ (blue) and
the square of the normalized residual intensity $I_{r}^{2}$ (black);
the covering factor $C_{\rm v}$; the opacity expressed as
$e^{-\tau_{\rm v}}$. Horizontal dashed lines indicates the range [0,1]
over which these quantities have a physical meaning. Unphysical values
of $C_{\rm v}$ and $e^{-\tau_{\rm v}}$ have been cut to a maximum of
1.1 and a minimum of -0.1}
\label{fig:cova}
\end{figure}

\begin{figure}
\resizebox{\hsize}{!}{\includegraphics*{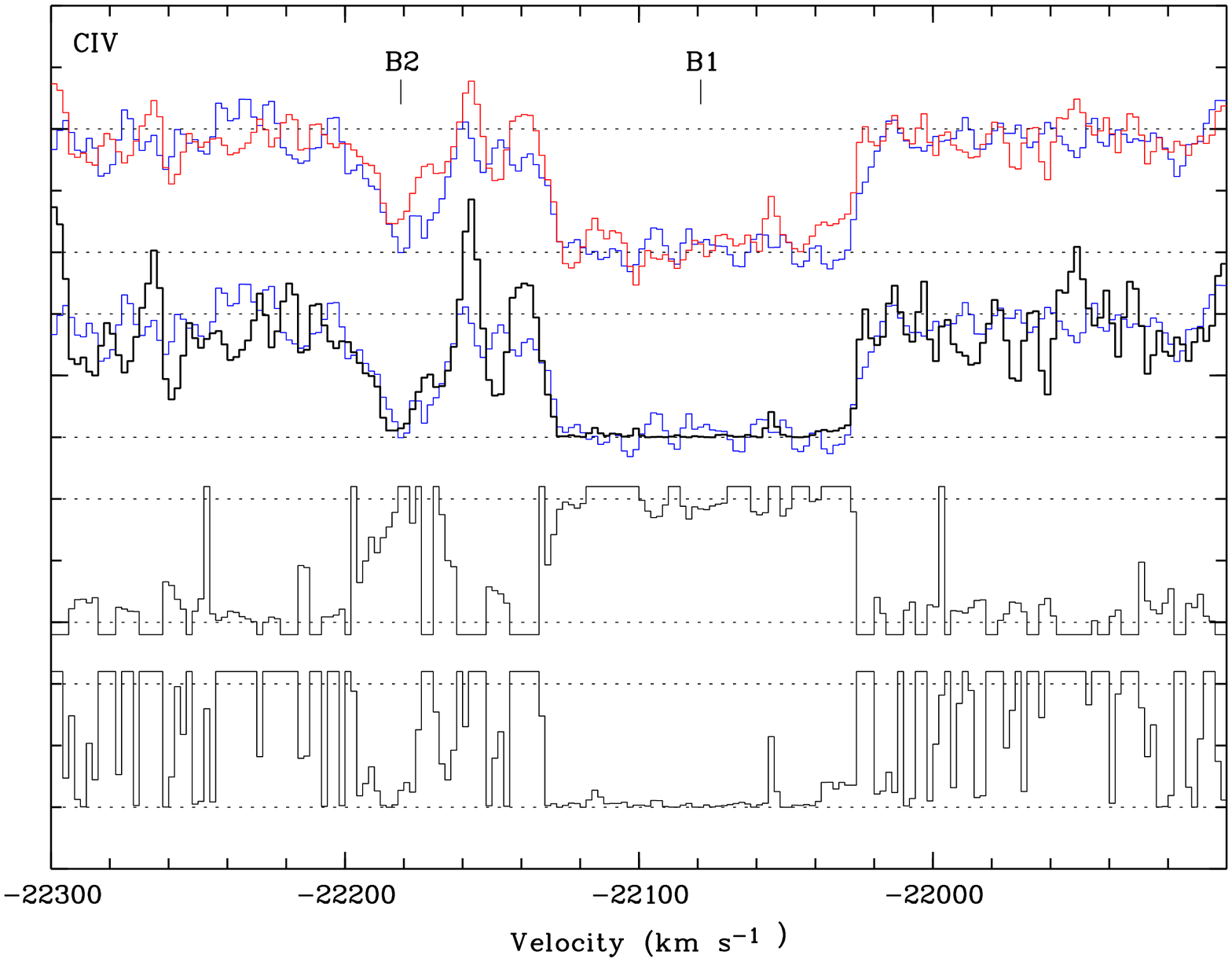}}
\resizebox{\hsize}{!}{\includegraphics*{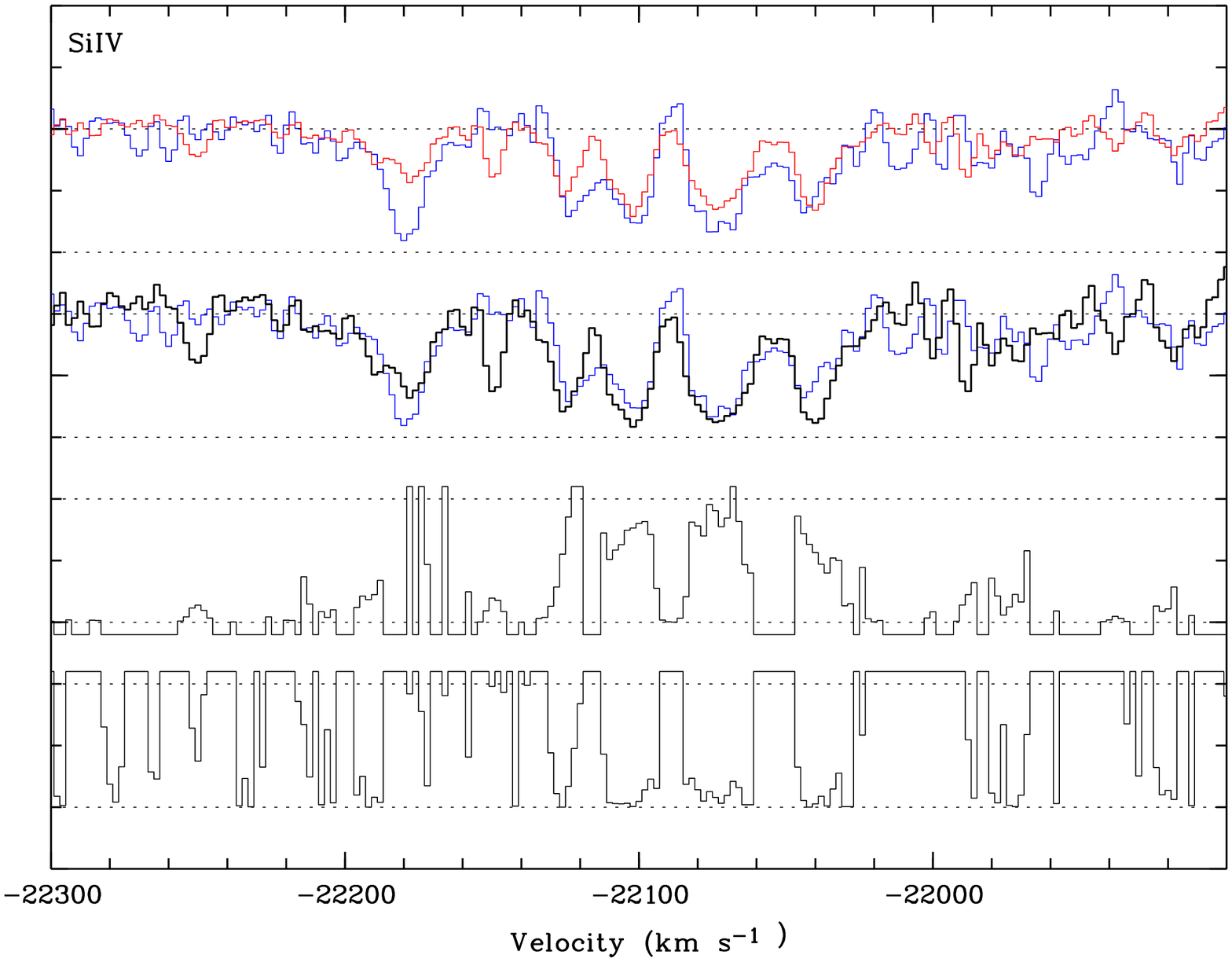}}
\caption{Same as Fig.\ref{fig:cova} but for the high-velocity
 absorption system B}
\label{fig:covb}
\end{figure}

Results for the low-velocity absorption system A are given in
Fig.~\ref{fig:cova}. Although some parts of the line profiles are
noisy or saturated, a solution to Eqs. (1) - (2) is found for most data
points in the absorption profiles, apart from a few relatively narrow
``spikes'' due to photon noise, incorrect background subtraction, or
contamination by other lines. These results are not very sensitive to
small modifications to the adopted local continuum.

A clear trend is observed especially in the lowest velocity component
A1: the line profiles are mostly determined by the velocity-dependent
covering factor while the absorption is completely saturated;
remarkably $e^{-\tau_{\rm v}} \simeq 0$ all throughout the
profiles. This complete saturation prevents the determination of
column densities. For \mgii\ and \siiv\ it is quite clear that the
normalized residual intensites $I_{r}$ and $I_{b}$ are identical
within the uncertainties.  Since these intensities were normalized to
a local continuum including different amounts of broad emission at the
wavelength of each doublet, the absorption must cover part of the
broad emission region, justifying our hypothesis a posteriori.
Components A2 and A3 in \siiv\ and \nv\ do follow the same trend,
although a spike obviously contaminates the \nv\ profile between
components A2 and A3.

The covering factor is also dependent on the ionization.  Apart from
the fact that the components A2 and A3 are not seen in \mgii , the
covering factor of the \mgii\ component A1 reaches a maximum value of
only $\sim$ 0.5, smoothly decreasing to lower and higher velocities.
For \siiv\ and \nv\ the derived covering factor has a broader profile
with a maximum value reaching nearly complete covering. No significant
difference is seen between \siiv\ and \nv .

\subsubsection{The high-velocity absorption system}

Results for the high-velocity narrow absorption system B are given in
Fig.~\ref{fig:covb}. \nv\ data are not illustrated due to the poor
signal to noise (Fig.~\ref{fig:vel}).  Moreover \nv\ in the
higher velocity component B2 is contaminated by intervening Ly$\alpha$
absorption, and nothing clear can be derived about its nature.

The \civ\ B1 component appears black and saturated at all velocities
indicating that it must fully cover the continuum emitting region
(there is no longer emission at these velocities).

More interesting is the covering factor derived for \siiv .  Although
the data are noisier than in system A, $I_{r}^{2} < I_{b}$ for the
four absorption sub-troughs seen in the B1 component, corresponding to
a velocity-dependent covering factor varying between 0.9 and 0.5.
This indication of partial covering supports the hypothesis that this
high-velocity component is intrinsic to the QSO.  Detection of
variability would nevertheless be useful to confirm this result.

Confirming the intrinsic nature of this high-velocity narrow component
would be interesting: high-velocity NALs are not uncommon but they are
rarely seen in BAL QSOs (Hamann et al.  \cite{ham1997a}).

\subsection{Absorption from excited levels: electron density and distance 
to the absorbers}

\begin{figure}[t]
\resizebox{\hsize}{!}{\includegraphics*{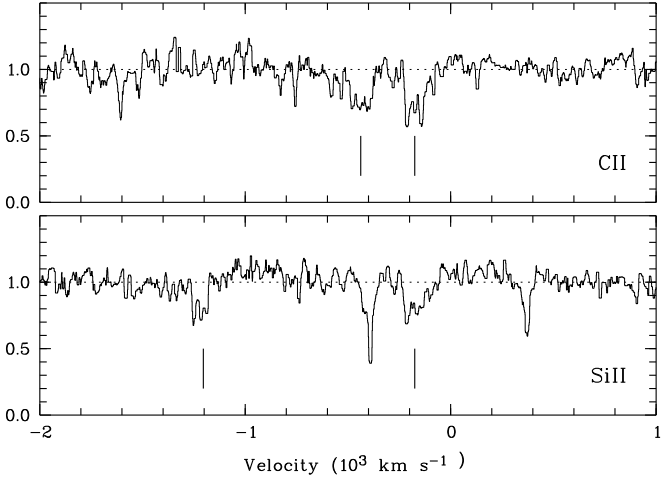}}
\caption{Part of the spectrum illustrating the \cii\
$\lambda$1334.53, \ciie\ $\lambda$1335.70, \siii\ $\lambda$1260.42,
and \siiie\ $\lambda$1264.76 transitions in system A1
(cf. Fig. \ref{fig:vel}). In order to emphasize these features, the
spectra are divided by a local continuum, and smoothed using median
filtering on 20 km s$^{-1}$ boxes. The vertical marks indicate
the velocities predicted for these absorption lines at $z = 2.29071$ (A1).}
\label{fig:velz}
\end{figure}

Excited-state narrow absorption lines of \ciie\ $\lambda$1335.7
and \siiie\ $\lambda$1264.8 are clearly detected in the component A1
in the spectrum of \sd\ (Fig.~\ref{fig:velz}). These features arise
from ground multiplets that behave approximately as two-level atoms,
the level populations being controlled by collisional processes and
radiative decays (Bahcall \& Wolf \cite{bah1968}, Morris et
al. \cite{mor1986}, Osterbrock \cite{ost1989}).  The strength of the
absorption lines from the excited fine-structure levels \ciie\
$\lambda$1335.7 and \siiie\ $\lambda$1264.8 can be directly compared
to the resonance transitions \cii\ $\lambda$1334.5 and \siii\
$\lambda$1260.4 in order to estimate the electron density needed to
populate the upper level.  Note that the very close transitions
\siiie\ $\lambda$1264.74 and \siiie\ $\lambda$1265.00 --as well as
\ciie\ $\lambda$1335.66 and \ciie\ $\lambda$1335.71-- are considered as
a single transition at the $gf$-weighted wavelength, keeping in mind
that one of the transitions is much stronger.  Also note that the weaker
transitions \siii\ $\lambda$1304.37 and \siiie\ $\lambda$1309.27 are
not detected, despite of confusion with \siiv\ from absorption system
B and an instrumental artifact (seen in all UVES spectra from the
observing run), respectively.

The population ratio of the upper excited level 2 to the lower 
resonance level 1 may be written (Osterbrock \cite{ost1989})
\begin{equation}
\frac{N_{\rm 2}}{N_{\rm 1}} = n_{e} \frac{q_{12}}{A_{21}} 
     (1+ n_{e} \frac{q_{21}}{A_{21}})^{-1} ,
\end{equation}
assuming equilibrium between collisional excitation, collisional
de-excitation and radiative de-excitation\footnote{In principle,
excitation by infrared radiation is also possible. However, for this
excitation mechanism to be significant, very low electron densities
are required, hence extremely large distances of the low-ionization
clouds from the photoionization source. Even if the infrared source is
extended, this requires infrared flux densities much larger than
expected for a typical quasar, and \sd , undetected by IRAS, is not an
infrared-luminous quasar.}.
$A_{21}$ is the radiative de-excitation rate; $n_e$ is the electron
density; $q_{21}$ = $8.629\,10^{-6} \, T^{-1/2}\, g_{2}^{-1} \,
\Omega(1,2)$ and $q_{12}$ = $q_{21} \, g_{2} \, g_{1}^{-1} \,
e^{-E_{12}/kT}$; $g_1$ = 2 and $g_2$ = 4 are the statistical weights
of levels 1 and 2; $E_{12}$ is the difference of energy between levels
1 and 2; $T$ is the electron temperature; and $\Omega(1,2)$ is the
collision strength, which is only slightly dependent on the
temperature (e.g. Hayes \& Nussbaumer \cite{hay1984}). For \siiie\
$\lambda$1264.8, $A_{21}$ = 2.13 10$^{-4}$ s$^{-1}$, $\Omega(1,2)$ =
5.58, and $E_{12}$ = 3.56 10$^{-2}$ eV (corresponding to the
fine-structure transition [\siii] at $\lambda$~34.8 $\mu$m); for
\ciie\ $\lambda$1335.7, $A_{21}$ = 2.29 10$^{-6}$ s$^{-1}$,
$\Omega(1,2)$ = 2.90, and $E_{12}$ = 7.86 10$^{-3}$ eV (corresponding
to the fine-structure transition [\cii] at $\lambda$~158 $\mu$m) (data
from Osterbrock 1989, and from The Atomic Line List v2.04).

\begin{figure}
\resizebox{\hsize}{!}{\includegraphics*{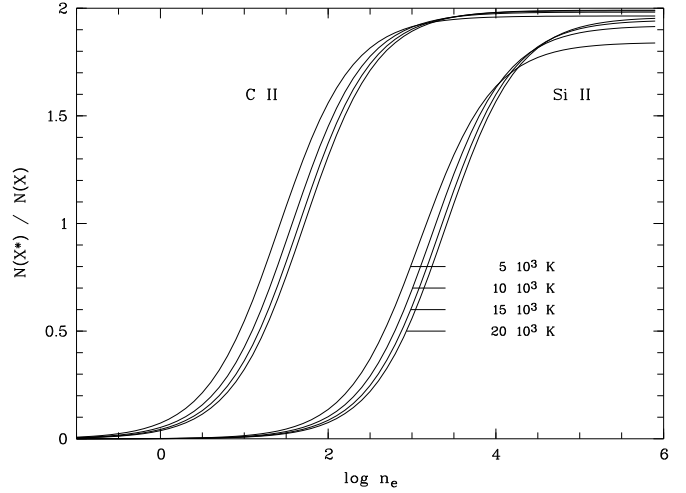}}
\caption{\cii\ and \siii\ column density ratios $N(X^{\star}) / N(X)$
as a function of the electron density $n_e$ in cm$^{-3}$. The \siii\
curves are labeled with the electron temperature; the \cii\ curves
follow the same trend.}
\label{fig:colli}
\end{figure}

Eq.~(3) has been used to compute the \cii\ and \siii\ column density
ratios for several values of the electron density and
temperature. Results are illustrated in Fig.~\ref{fig:colli} and can
be used as a diagnostic.  It is interesting to note that \siii\ and
\cii\ are sensitive to different --although overlapping--
electron density ranges.

Looking at Fig.~\ref{fig:velz} we can see that \ciie\
$\lambda$1335.7 is stronger than \cii\ $\lambda$1334.5 while \siiie\
$\lambda$1264.8 is approximately equal to \siii\ $\lambda$1260.4.
Although the signal to noise is not optimal and uncertainties on the
location of the continuum are large, we estimate $N$(\ciie)/$N$(\cii)
$\simeq$ 1.6$\pm$0.3 and $N$(\siiie)/$N$(\siii) $\simeq$ 1.1$\pm$0.35.
These values were first obtained by integrating the optical depths
over the excited- and ground-state line profiles, and then computing
their ratio.  Also, in order to ensure that the same parts of the
profiles are considered, the optical depth ratio was computed as a
function of the velocity, and then the mean ratio evaluated. Both
methods give similar results.  The uncertainties are estimated by
considering the ratios obtained with these two methods, as well as
various binning values and locations of the underlying continuum.

According to Fig.~\ref{fig:colli}, the measured values roughly agree
within the uncertainties to indicate an electron density $\log n_{e}
\approx$ 3, the value derived from \cii\ being smaller than
the one derived from \siii . The overall agreement between the \cii\
and \siii\ behaviors supports the assumptions underlying Eq.~(3).
However, these column density ratios have been computed assuming
complete covering ($C_{\rm v} = 1$), which is most probably not
true. When there is partial covering, unabsorbed flux is added to the
line profiles with the result that intensity differences between two
lines are attenuated. Correcting for partial covering will restore
these differences. Taking partial covering into account could then
increase $N$(\ciie)/$N$(\cii) and decrease $N$(\siiie)/$N$(\siii),
which would result in a better agreement between the electron
densities derived from both species. In the following we
conservatively adopt $2.2 \le \log n_{e} \le 3.4$, the lower and
higher values corresponding to the values obtained from \cii\ and
\siii\ respectively, with $T = 10^4$~K.

With the reasonable assumption that the gas is in photoionization
equilibrium with the quasar radiation field, the electron density can be
combined with the ionization parameter $U$ to estimate the distance $r$
between the absorber and the quasar.  Hamann et al. (\cite{ham2001}) give
\begin{equation}
r = \left( \frac{L_{\rm LL}}{4 \pi c h  (-\alpha)  n_{\rm H} U} \right) ^{1/2}
\end{equation}
where $L_{\rm LL}$ is the luminosity density at the Lyman limit,
$n_{\rm H}$ is the total hydrogen density, and $\alpha$ the power-law
spectral coefficient in the Lyman continuum.  The ionization parameter
$U$ is constrained by the simultaneous presence of \mgii\ and \nv\ in
the narrow component A1, which requires $-2.8 \lesssim \log U \lesssim
-2.5$ for ionization fractions $>1\%$ for both ions, according to the
calculations of Hamann \& Ferland (\cite{ham1999}).  We adopt $\log U
= -2.8$ for ease of comparison with Hamann et al.  (\cite{ham2001}).
We estimate $L_{\rm LL} \approx$ 2.5 10$^{30}$ ergs s$^{-1}$ Hz$^{-1}$
by extrapolating the flux measured in the low-resolution SDSS spectrum
(assuming a cosmology with $H_0$ = 70 km s$^{-1}$ Mpc$^{-1}$,
$\Omega_M = 0.3$, and $\Omega_{\Lambda} = 0.7$, Pen et
al. \cite{pen1999}).  Adopting $n_{\rm H} \approx n_e$, and $\alpha =
-1.6$, again for ease of comparison with Hamann et
al. (\cite{ham2001}), we find a distance 9 kpc $\le r \le$ 37
kpc.  This range of values reflects only the uncertainties on $n_e$; the
distance would be 30\% smaller if we had used $\log U = -2.5$.  We
emphasize that this distance is only a rough estimate, and in
particular relies upon the assumption of a uniformly dense outflow, as
discussed in the next section.

\section{Discussion and Conclusions} 
 \label{sec:Dis}

One of the main results of our study is that the narrow absorption
components seen at the low-velocity end of the BAL troughs are
completely saturated, their profiles being dominated by the covering
factor. Velocity and ionization dependent partial covering is often
seen in NALs (Barlow \& Sargent \cite{bar1997a}, Hamann et
al. \cite{ham2001}), and has also been reported for mini-BAL and BAL
QSOs (Arav et al. \cite{ara1999b}, Srianand \& Petitjean
\cite{sri2000}, de Kool et al. \cite{dek2001}, \cite{dek2002b}, Hall et
al. \cite{hal2003}).  The partial covering proves the intrinsic nature
of the absorbers.

However, our most intriguing result is the distance between the
ionizing source and the absorber that we derived on the basis of the
excited lines.  BALs are usually thought to be formed much closer to
the central engine, at distances $\sim$ 0.1 -- 1 pc, roughly four to
five orders of magnitude smaller than the distance of $\sim$ 20
kpc we have measured. Our estimate is clearly uncertain, but the
errors cannot explain such a large difference.

Note that the \ciie\ and \siiie\ must be related to the other
absorption features.  The correspondence in velocity of the high- and
low-ionization features in component A1 (Figs.~\ref{fig:vel} and
\ref{fig:velz}) clearly indicate that all species must form at roughly
the same location.  Also, the continuity of the opacity and covering
factor properties towards the higher velocity components A2 and A3
(Fig. \ref{fig:cova}) suggests a common formation for these components
too.  The components of the absorbing system A are definitely part of
the BAL outflow because of their association with the high-ionization
features seen in the wider \civ\ BAL (cf. the SDSS
low-resolution spectrum in Fig.~\ref{fig:sdss}).

It is therefore likely that the BAL formation region (BALR) in \sd\
--- at least for component A --- is at much higher distance than
commonly thought.  In fact, small BALR distances are based more on
theoretical considerations than on direct observational evidence.
Here we summarize the few direct estimates that have been made of the
distances to confirmed intrinsic outflows; all distances have been
converted to our cosmology (Sect. 4.2).

Two NALs that have undergone time variability have thereby confirmed
themselves as intrinsic and yielded upper limits to their distances in
the kiloparsec range: $<$ 1 kpc in UM~675 (Hamann et
al. \cite{ham1997b}), and $<$ 2 kpc in QSO 2343+125 (Hamann et
al. \cite{ham1997c}).

Large distances have been inferred from the presence of \ciie\ or
\siiie\ absorption in two other NALs which are known to be intrinsic
absorbers due to partial covering: 20 kpc for 3C~191 (Hamann et al.
\cite{ham2001})\footnote{Note that 3C 191 is a radio-loud quasar,
whereas \sd\ is radio-quiet by virtue of its non-detection in the
FIRST survey (Becker et al. \cite{bwh1995}). Absorbers apparently can
be found at large distances from the ionizing source in both RLQs and
RQQs.},  and $>$ 570 pc for APM 08279+5255 (Srianand \& Petitjean
\cite{sri2000}).

A related but distinct technique is the use of numerous \ion{Fe}{ii}
absorption lines to estimate absorber distances.  Using this method,
$r$ $\sim$ 310 pc has been measured for the low-velocity system in FIRST
J084044.5+363328, whereas $r$ $\sim$ 1.35 pc is found for the
high-velocity system in the same object using different diagnostic (de
Kool et al. \cite{dek2002b}).  However, analysis of the similar NAL in
FIRST J121442.3+280329 yielded a small distance range $\sim$ 1 -- 30 pc
for the entire outflow (de Kool et al. \cite{dek2002a}), similar to
that found for LBQS 0059-2735 (Wampler et al. \cite{wam1995}). The
largest distance found by this method has been for the low-ionization
BAL in FIRST J104459.6+365605 (de Kool et al.  \cite{dek2001}).  On
the basis of excited \ion{Fe}{ii} lines and velocity correspondence
between \ion{Fe}{ii} and \mgii\ lines, $r$ $\sim$ 630 pc was found.  We
emphasize that -- again based on velocity correspondence -- our high
distance estimate in \sd\ also applies to the high-ionization
absorbers and not only to the low-ionization ones, as found in FIRST
J104459.6+365605.

However, Everett et al. (\cite{eka2002}) model a multiphase outflow at
only $r$ $\sim$ 4 pc that can reproduce the observations of FIRST
J104459.6+365605, including the density measurements used by de Kool
et al. to infer a much larger distance.  In their model the intrinsic
outflow does not have a constant density, but instead consists of a
relatively low-density, high-ionization wind with embedded
higher-density, lower-ionization clouds.  Absorption in the
high-ionization wind modifies the spectrum seen by the outer region of
the wind and the low-ionization clouds, effectively reducing the value
of $L_{\rm LL}$ in Eq.~(4) and thereby decreasing the inferred distance.
The ionization and density in the wind decrease with increasing
distance until species such as \ion{Fe}{ii} and \mgii\ are present at
the required $n_e$.  Denser clouds embedded in the flow at that
distance are invoked to produce \ion{Mg}{i} absorption at the same
velocities as \ion{Fe}{ii} and \mgii .

Everett et al. claim that a multiphase outflow at a small distance
from the ionizing source could also explain the absorption in 3C 191.
Hamann et al. do in fact state that the outflow in 3C 191 must span a
range of densities (or distances) to explain the presence of \nv\ and
\ion{Mg}{i} at the same velocities.  A range of densities could also
be present in \sd\ but is not required, because \ion{Mg}{i} is not
detected.  In any case, the density diagnostics used to infer large
absorber distances in \sd\ and 3C 191 are different than those used in
FIRST J104459.6+365605.  Therefore, detailed modeling should be done
to determine whether or not a multiphase outflow at small distances
can indeed reproduce the observed densities, column densities and
velocity structure of all ions observed in these objects.  A further
observational test of a multiphase model for \sd\ and 3C 191 could
come from spectra extending shortward of Ly$\alpha$ past the Lyman
limit, directly measuring $L_{\rm LL}$ and constraining any
modification of the spectrum by the inner wind.  It would also be
worth monitoring objects with large inferred distances, to search for
time variable NALs which could require much smaller distances.

If they are confirmed, large distances from the continuum source and
the broad emission line region may be difficult to reconcile with the
evidence of velocity and ionization dependent partial covering in
these absorbers.  On the one hand, partial covering implies only that
the projected size scale of the absorber is less than or comparable to
that of the emitting region.  {\em That inferred size for the absorber
is independent of the distance from the source,} as long as the
distance is small compared to the angular diameter distance to the
quasar.  Partial covering by outflows at kpc-scale distances is
therefore possible, at least in principle.  However, it is difficult
to understand how such distant outflows could be common, for several
reasons.  The usual problems of cloud (or density inhomogeneity)
survival and confinement (e.g. Hamann et al. \cite{ham2001}, de Kool
et al.  \cite{dek2001}) are exacerbated by the requirement that such
structures must survive for the time needed to reach such large
distances.  Also, if such outflows are seen in $\sim$10\% of quasars,
they cover somewhere between $\sim$10\% of the unobscured lines of
sight around all quasars and all such sightlines in $\sim$10\% of
quasars at distances of $\sim$10~kpc, implying outflows of very large
masses ($\sim 10^8-10^9 M_{\sun}$; Hamann et al. \cite{ham2001}).

It is worth noting that the observed partial covering may in fact be
due to additional emission from an extended region of scatterers
comparable in size to the BALR.  Such a region has already been
suggested to explain spectropolarimetric measurements (Cohen et
al. \cite{coh1995}, Goodrich \& Miller \cite{goo1995}) but its size
and location have never been directly measured.  Resonance scattering
in a roughly axially symmetric BALR could also explain abnormal
doublet line ratios usually interpreted as partial covering
(e.g. Branch et al. \cite{bra2002}).  Spectropolarimetry could provide
some tests of these hypotheses.

Whether our results are generic to BAL QSOs is not clear. The objects
for which BALR sizes have been measured so far show absorption
somewhat intermediate between mini-BALs and BALs, rather than the very
wide and deep troughs often seen in BAL QSOs. Perhaps these
intermediate objects have different outflows, or represent an older
evolutionary stage of the BAL QSO phenomenon when most of the material
has dissipated, preferentially leaving behind denser clumps.

Clearly these results are puzzling and raise many questions.  To
determine the range of distances spanned by intrinsic outflows will
require detailed study of other, similar objects with \ciie\ and
\siiie\ absorption, including photoionization modeling to determine if
multiphase models can explain such outflows.

 \begin{acknowledgements}
       We acknowledge use of data from the Atomic Line List v2.04
       (http://www.pa.uky.edu/$\sim$peter/atomic/).
	PBH acknowledges support from Fundaci\'on Andes.
	Funding for the creation and distribution of the SDSS Archive has been
	provided by the Alfred P. Sloan Foundation, the Participating
	Institutions, the National Aeronautics and Space Administration, the
	National Science Foundation, the U.S. Department of Energy, the
	Japanese Monbukagakusho, and the Max Planck Society. The SDSS Web site
	is http://www.sdss.org/.  The SDSS is managed by the Astrophysical
	Research Consortium (ARC) for the Participating Institutions. The
	Participating Institutions are The University of Chicago, Fermilab, the
	Institute for Advanced Study, the Japan Participation Group, The Johns
	Hopkins University, Los Alamos National Laboratory, the
	Max-Planck-Institute for Astronomy (MPIA), the Max-Planck-Institute for
	Astrophysics (MPA), New Mexico State University, University of
	Pittsburgh, Princeton University, the United States Naval Observatory,
	and the University of Washington.
 \end{acknowledgements}

\end{document}